\documentclass{eas}

\usepackage{graphicx}
\usepackage{hyperref}
\hypersetup{colorlinks=true,linkcolor=blue,citecolor=blue,urlcolor=blue}
\newcommand{\degree}{\ensuremath{^\circ}}

\begin{document}

	\title{Exploring the water and carbon monoxide shell around Betelgeuse with VLTI/AMBER}\thanks{Based on AMBER observations made with ESO Telescopes at the Paranal Observatory under programmes ID 086.D-0351 and 286.D-5036}
	\runningtitle{$H_2 O$ and $CO$ around Betelgeuse analysed with VLTI/AMBER}
	\author{M. Montarg\`es}
	\address{LESIA, Observatoire de Paris, CNRS, UPMC, Universit\'e Paris-Diderot, 5 place Jules Janssen, 92195 Meudon, France}
	\author{P. Kervella}
	\sameaddress{1}
	\author{G. Perrin}
	\sameaddress{1}
	\author{K. Ohnaka}
	\address{Max-Planck-Institut f\"ur Radioastronomie, Auf dem H\"ugel 69, 53121 Bonn, Germany}
	
	\begin{abstract}
		We present the results of the analysis of our recent interferometric observations of Betelgeuse, using the AMBER instrument of the VLTI. Using the medium spectral resolution mode ($R \sim 1500$) we detected the presence of the water vapour and carbon monoxide (CO) molecules in the H and K bands. We also derived the photospheric angular diameter in the continuum. By analysing the depth of the molecular lines and the interferometric visibilities, we derived the column densities of the molecules, as well as the temperature and the size of the corresponding regions in the atmosphere of Betelgeuse (the MOLsphere) using a single shell model around the photosphere. Our results confirm the findings by Perrin \etal\ (\cite{Perrin2004}) and Ohnaka \etal\ (\cite{Ohnaka2011}) that the H$_2$O and CO molecules are distributed around Betelgeuse in a MOLsphere extending to approximately 1.3 times the star's photospheric radius.
	\end{abstract}
	
	\maketitle
	
	\section{Introduction}\label{Sect_Introduction}

		Betelgeuse was observed in January and February 2011 with the ESO Very Large Telescope Interferometer (VLTI) using the Astronomical Multi-BEam combineR, AMBER (the instrument is described in Petrov \etal\ \cite{Petrov_AMBER}) in the $H$-band from 1.45 to 1.80 $\mu$m and in the $K$-band from 2.10 to 2.45 $\mu$m with the E0-G0-H0, E0-G0-I1 and G0-H0-I1 triplets. The CO and water vapor absorption lines in the $K$ band were investigated using the medium spectral resolution of the instrument ($R \sim 1500$).

	\section{Data reduction}\label{Sect_Data_Reduction}
	
		The AMBER data reduction package version 3.0.3 also known as \emph{amdlib} was used to obtain the calibrated interferometric observables (Tatulli \etal\ \cite{Tatulli_AMBER}) : the three visibilities associated to the three baselines are directly the amplitude of the object's Fourier Transform. It is not possible to measure directly the phase but the calibrated data contain three differential phases (DP) which are sensitive to the photocenter shift in a spectral line with respect to the continuum. The closure phase (CP) is the sum of three phases along the three baselines that form a triangle : $\phi_{CP} = \phi_{12} + \phi_{23} + \phi_{31}$. This quantity is independent from the atmospheric perturbations.
		
		The wavelength calibration was performed using telluric lines in the spectrum of the interferometric calibrator HR-1543 (spectral type F6V). The error bars in the visibilities and closure phases of Betelgeuse were underestimated, probably due to the low visibility of the fringes. We had to split each data file into five subsets to get a new estimation of these errors.  We also noticed that some datasets were showing unusually low visibilities : we conclude it was caused by lock loss of the fringe tracker FINITO. These corrupted data sets were discarded.

	\section{Results}\label{Sect_Results}
	
		\subsection{Continuum analysis}\label{SubSect_Continuum}
		
\begin{figure}
			\centering
			\includegraphics[width=5cm]{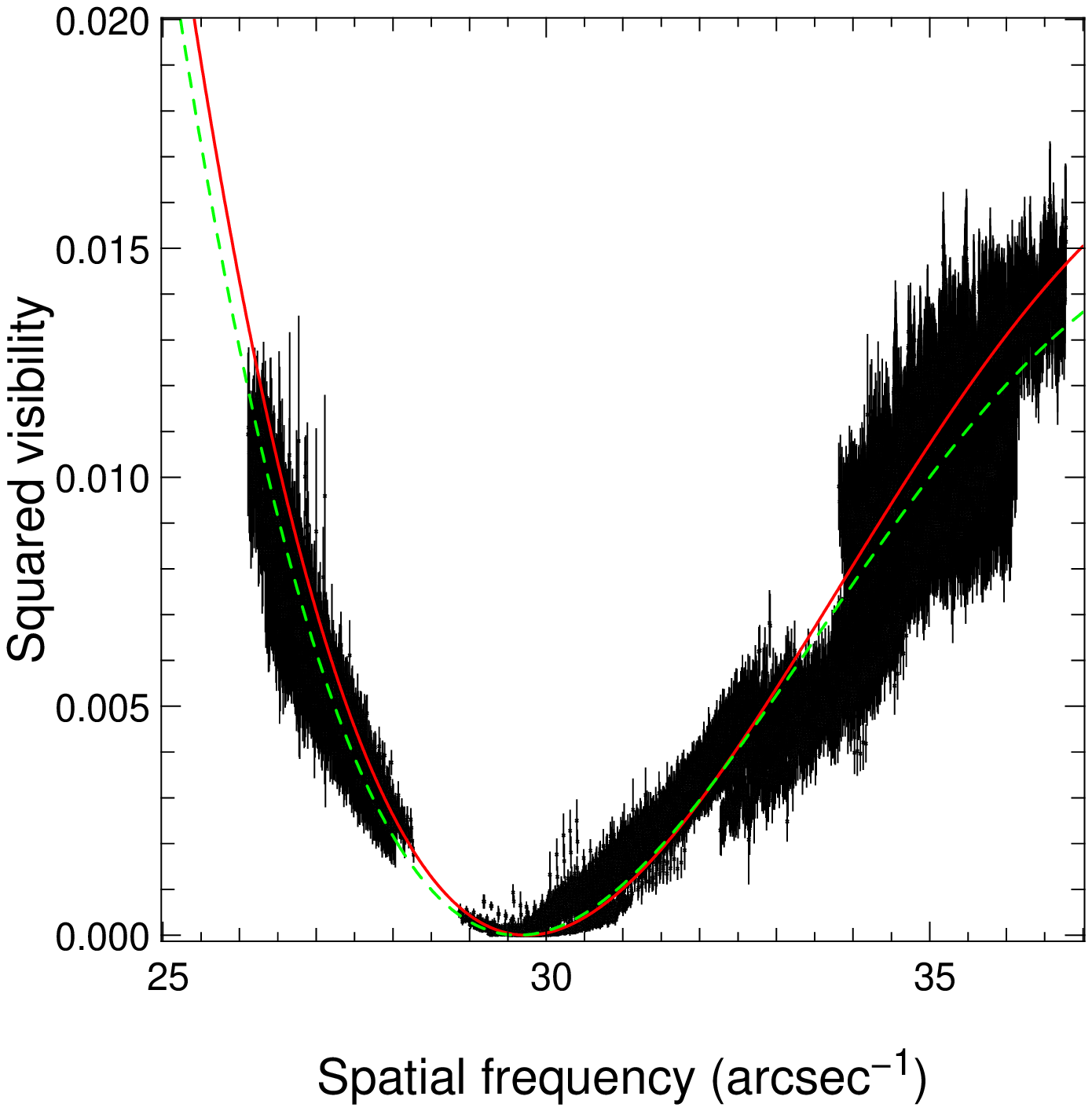}
			\qquad
			\includegraphics[width=5cm]{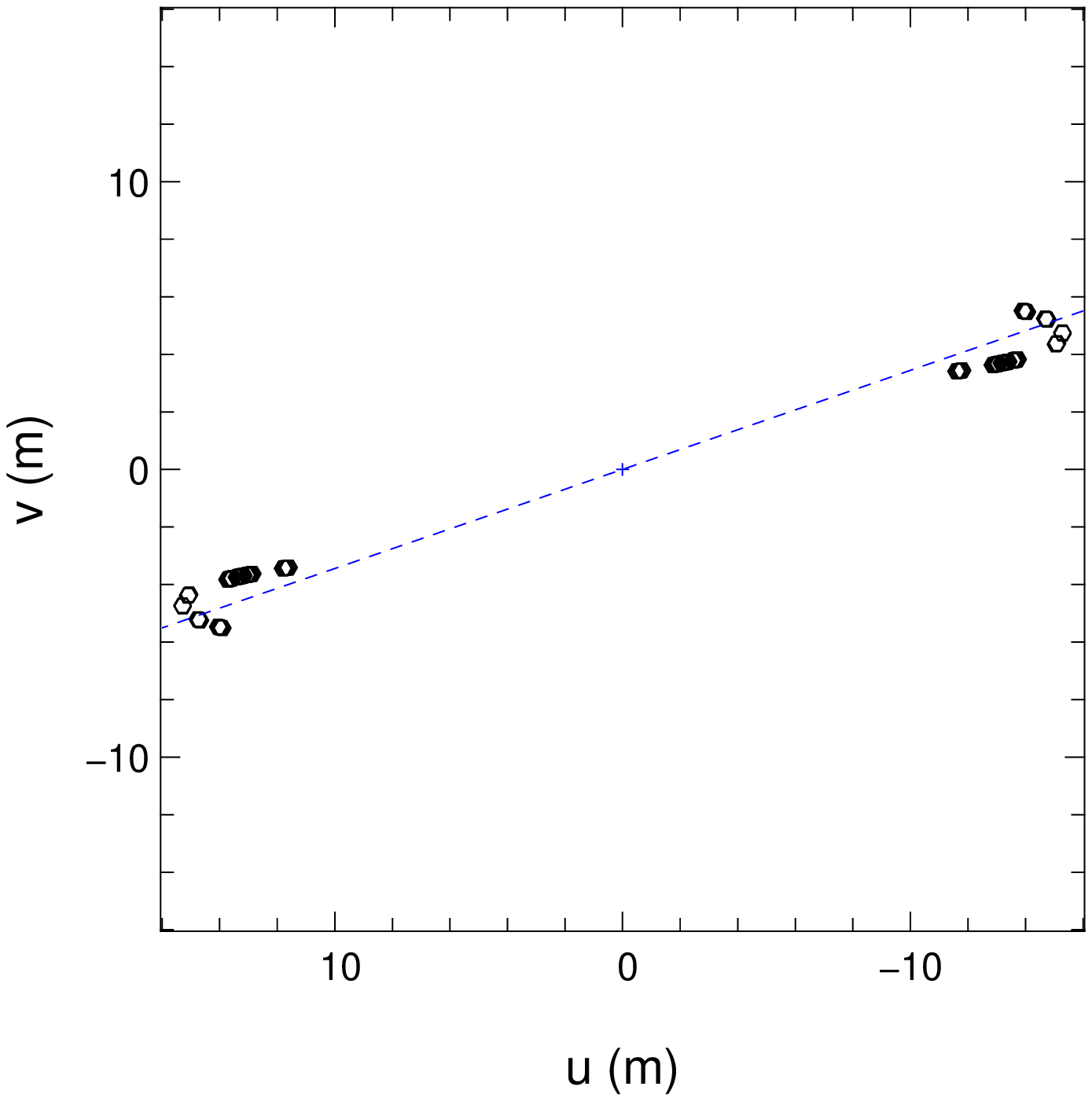}
			\caption{\textit{Left: }Fit of the continuum squared visibilities. The black points represent the data. The red continous line is the best fit uniform disk model and the green dashed line is the best fit limb-darkened disk model. \textit{Right: }$(u,v)$ coverage of our AMBER data for the first and second lobes. The blue dashed line represents the average position angle of 73$\degree$.}
			\label{Fig_Continuum}
		\end{figure}

			We used two different models to fit the continuum visibilities between 2.1 and 2.245 $\mu$m : a uniform disk (UD) and a limb-darkened disk (LDD) using a power law ($I = I_0 \mu^\alpha$) according to Hestroffer (\cite{Hestroffer1997}). We only fitted the data from the first and second lobes in order to avoid contributions  from small scale structures at higher spatial frequencies. We used 8752 visibility points over 31\,577, from which we derived an angular diameter of $41.01 \pm 0.08 \, \mathrm{mas}$ for the uniform disk model giving a reduced $\chi^2$ of 5.27. For the limb-darkened model we obtained an angular diameter of $42.28 \pm 0.09 \, \mathrm{mas}$ and a power law coefficient of $0.155 \pm 0.005$ with a reduced $\chi^2$ of 4.91. The result of these fits is presented in Fig. \ref{Fig_Continuum} together with the $(u,v)$ coverage in the first and second lobes. Our value of the angular diameter may not be representative of the average angular diameter of the star as we are only probing one direction of the $(u,v)$ plane.
 
			Our UD value is significantly lower than the previous ones found by Perrin \etal\ (\cite{Perrin2004}) of $43.26 \pm 0.04$ mas and Ohnaka \etal\ (\cite{Ohnaka2009} and \cite{Ohnaka2011}) of $42.05 \pm 0.05 \, \mathrm{mas}$ which was obtained for this later with VLTI/AMBER observations in the same $(u,v)$ direction. This could be caused by the relative decrease of the diameter shown in Ohnaka \etal\ (\cite{Ohnaka2011}) or by a change in the  flux distribution on the photosphere of the star. On the other hand, our LDD value agrees well with the $42.49 \pm 0.06 \, \mathrm{mas}$ of Ohnaka \etal\ (\cite{Ohnaka2011}).

		\subsection{H$_2$O and CO absorption lines : model fitting}\label{SubSect_ModelFitting}

\begin{figure}
			\centering
			\includegraphics[width=5cm]{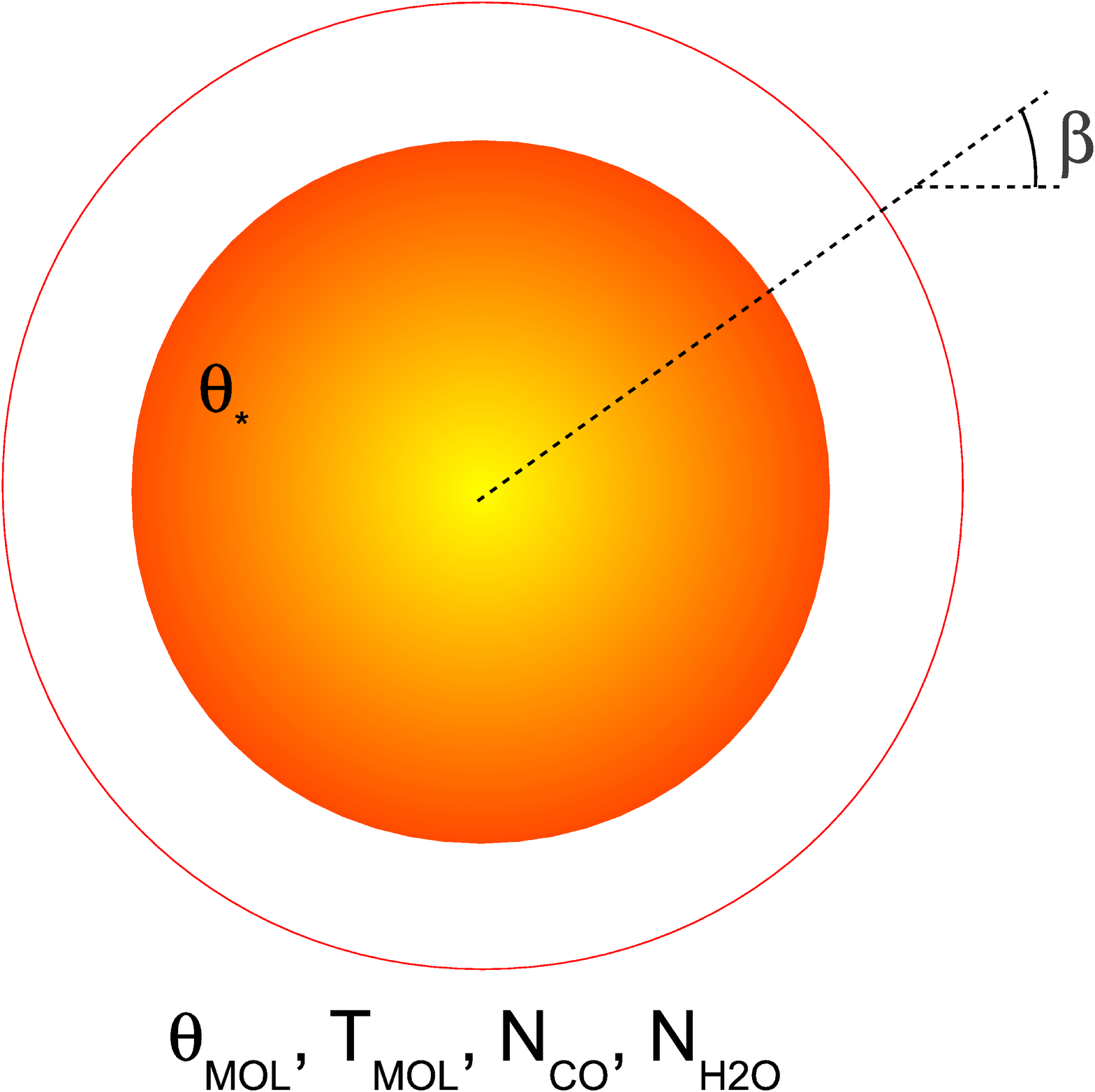}
			\qquad
			\includegraphics[width=5.3cm]{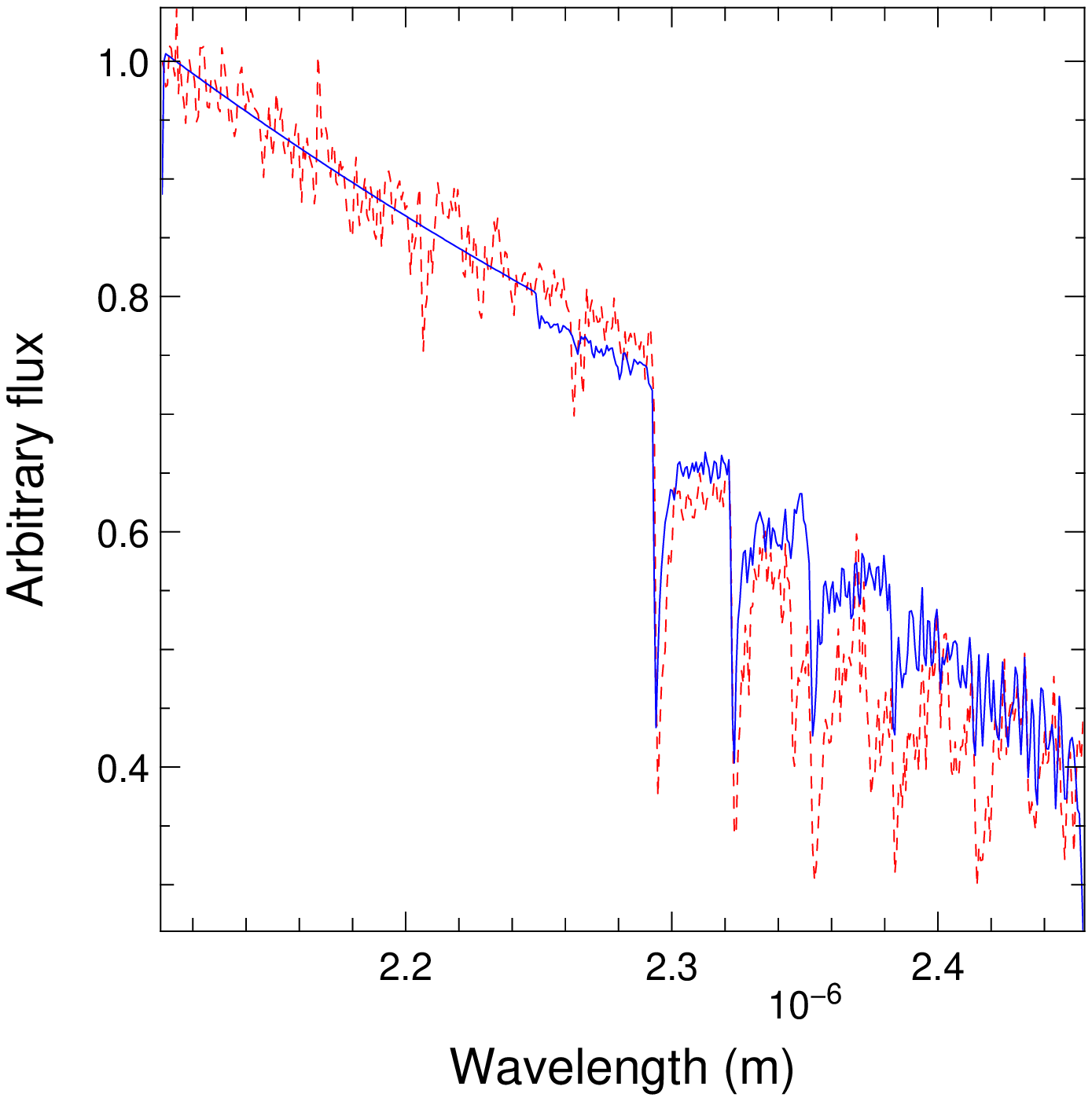}
			\caption{\textit{Left: }Illustration of the single layer model used to model the MOLsphere. \textit{Right: }The red dashed line is the spectrum obtained from the Betelgeuse AMBER data and the blue continuous line is the spectrum obtained from the single layer model. We used the best fit values from the equation \ref{Eq_Results}.}
			\label{Fig_Model}
		\end{figure}

			The wavelengths longer than 2.245 $\mu$m contain absorption lines from carbon monoxide (CO) and water vapor (H$_2$O) which are elements present in the MOLsphere described by Tsuji (\cite{Tsuji2000}) and Perrin \etal\ (\cite{Perrin2004}). In this paper we restrain ourselves to the first two $CO$ overtone band heads ($2.245 < \lambda < 2.348 \, \mu$m). It is clear from Fig. \ref{Fig_Model} (right panel) that the other bands are not well fitted when these first two are.\\
			
			We modeled them using a Kurucz atmosphere model\footnote{\url{http://kurucz.harvard.edu/}} (Castelli \& Kurucz \cite{Castelli2003} and Kurucz \cite{Kurucz2005}) and a single layer containing both CO and H$_2$O at the same temperature and distance from the photosphere and the corresponding absorptions were computed from the line list of Goorvitch (\cite{Goorvitch1994}) for CO and from Partridge \& Schwenke  (\cite{Partridge1997}) for H$_2$O. This model contains five parameters : the photospheric diameter $\theta_*$, the MOLsphere diameter $\theta_{\mathrm{MOL}}$, the MOLsphere temperature $T_{\mathrm{MOL}}$, the CO and H$_2$O column densities $N_{\mathrm{CO}}$ and $N_{{\mathrm{H}_2\mathrm{O}}}$.  Its analytical expression is given by :\\
				
				\begin{equation}
				\begin{array}{l}
					I_{N_{\mathrm{CO}},N_{{\mathrm{H}_2\mathrm{O}}}}(\lambda,\beta) = I_{\mathrm{Kurucz}} \ e^{-\tau(N_{\mathrm{CO}},N_{{\mathrm{H}_2\mathrm{O}}};\lambda) \over \cos(\beta)}\\ + B(\lambda,T_{\mathrm{MOL}})[1-e^{-\tau(N_{\mathrm{CO}},N_{{\mathrm{H}_2\mathrm{O}}};\lambda) \over \cos(\beta)}] 
				\end{array}
				\end{equation}
				 if $\sin(\beta)\leq {\theta_* \over \theta_{\mathrm{MOL}}}$ and otherwise :
				\begin{eqnarray}
				I_{N_{\mathrm{CO}},N_{{\mathrm{H}_2\mathrm{O}}}}(\lambda,\beta) = B(\lambda,T_{\mathrm{MOL}})[1-e^{-2\tau(N_{\mathrm{CO}},N_{{\mathrm{H}_2\mathrm{O}}};\lambda) \over \cos(\beta)}]
				\end{eqnarray}
				$B(\lambda,T)$ is the Planck function, $\beta$ is the angle between the line of sight and the center of the star and $\tau(N_{\mathrm{CO}},N_{{\mathrm{H}_2\mathrm{O}}};\lambda)$ is the MOLsphere opacity computed from the line lists of Goorvitch (\cite{Goorvitch1994}) and Partridge \& Schwenke (\cite{Partridge1997}). The squared visibility is then derived by computing the Hankel transform of this analytical expression :				
				\begin{equation}
					V_\lambda (x) = { \int_0^1 I(\lambda,r)J_0(rx)rdr \over \int_0^1 I(\lambda,r)rdr} 
				\end{equation}
			
				With $x = {\pi b \theta_* \over \lambda}$, $r = \sin(\beta)$ and $J_0$ the zeroth order Bessel function of the first kind.\\			
			
			An illustration of this model and an example of the corresponding spectrum of Betelgeuse model are given in Fig. \ref{Fig_Model}, together with our AMBER observation.\\
			
			The model contains five \textit{a priori} independent parameters but there are some correlations between $T_{\mathrm{MOL}}$ and $\theta_{\mathrm{MOL}}$ and between the two column densities. Because of this degeneracy, we had to develop a strategy to perform our fit. We fixed $\theta_*$ to the value obtained by the uniform disk fit of the continuum data. Then we computed $\chi^2$ maps of our data by fitting $N_{\mathrm{CO}}$ and then $N_{{\mathrm{H}_2\mathrm{O}}}$ at given $T_{\mathrm{MOL}}$ and $\theta_{\mathrm{MOL}}$.  :  
   				\begin{eqnarray}
      				\chi^2(T_{\mathrm{MOL}},\theta_{\mathrm{MOL}}) = \sum_{i=1}^N \left({V^2_i - V^2_{\mathrm{model}}(T_{\mathrm{MOL}},\theta_{\mathrm{MOL}}, N_{\mathrm{CO}}, N_{{\mathrm{H}_2\mathrm{O}}};S_i) \over \sigma_i}\right)^2 
   				\end{eqnarray} 

			where $S_i$ is the spatial frequency. $N_{\mathrm{CO}}$ and $N_{{\mathrm{H}_2\mathrm{O}}}$ are fitted on each cell of the grid $(T_{\mathrm{MOL}},\theta_{\mathrm{MOL}})$. However, there was also a correlation between $T_{\mathrm{MOL}}$ and the two column densities. We had to eliminate it by including the spectrum into our fit : this compensates for the lack of interferometric data in the band core caused by the band fringe fitting described in Sect. \ref{Sect_Data_Reduction}. 
			
\begin{figure}
			\centering
			\includegraphics[width=6cm]{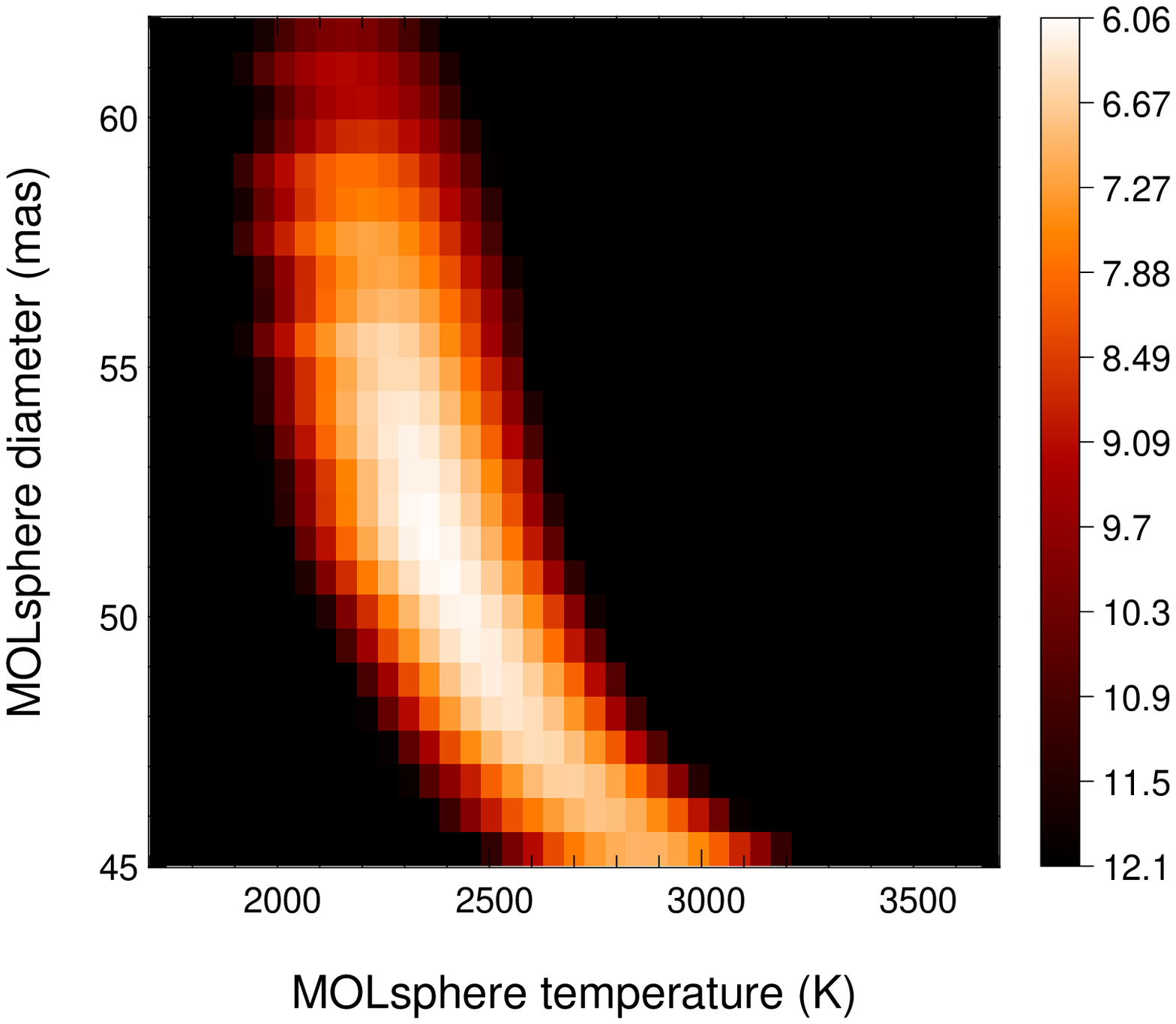}
			\quad
			\includegraphics[width=6cm]{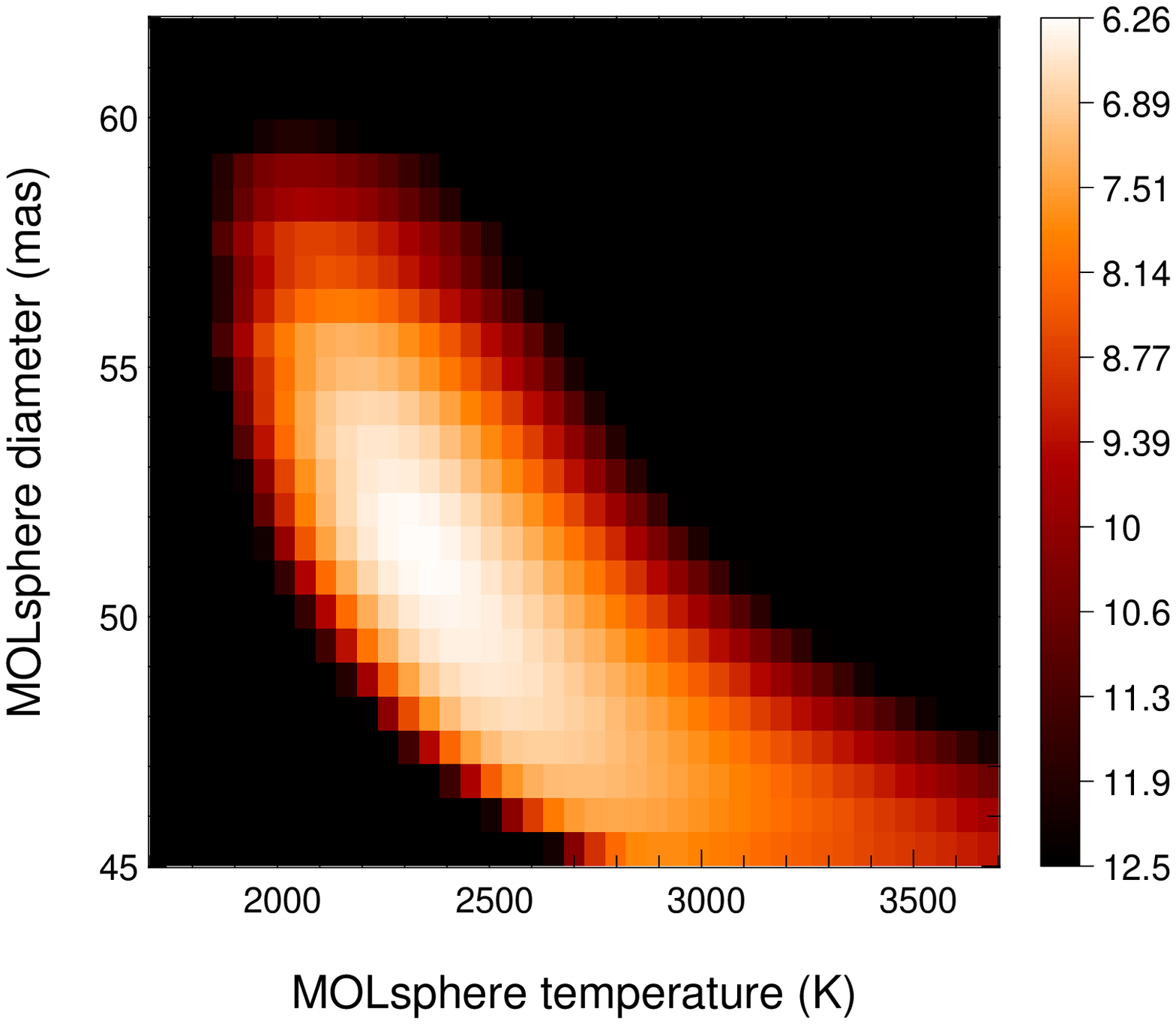}
			\caption{$\chi^2$ map of the single layer model. \textit{Left: }the CO column density is fitted on each cell of the grid for a constant $N_{{\mathrm{H}_2\mathrm{O}}} = 3.28 \times 10^{20} \, \mathrm{cm}^{-2}$. \textit{Right: }the H$_2$O column density is fitted on each cell of the grid for a constant $N_{\mathrm{CO}} = 1.53 \times 10^{21} \, \mathrm{cm}^{-2}$.}
			\label{Fig_Chi2Map}
		\end{figure}

			The $\chi^2$ maps are presented on Fig. \ref{Fig_Chi2Map} and we derived the following values for the parameters of the MOLsphere :
								
			\begin{equation}
			\begin{array}{l}
		   	T_{\mathrm{MOL}} = 2300 \pm 50 \, \mathrm{K}\\
		   	\theta_{\mathrm{MOL}} = 51.38 \pm 0.70 \, \mathrm{mas}\\
		   	N_{\mathrm{CO}} = (1.53 \pm 0.6) \times 10^{21} \, \mathrm{cm}^{-2}\\ 
			N_{{\mathrm{H}_2\mathrm{O}}} = (3.28 \pm 0.7) \times 10^{20} \, \mathrm{cm}^{-2}
		   	\end{array}
		   	\label{Eq_Results}
		  	\end{equation}	
		  	
			Our two $\chi^2$ maps for carbon monoxide and water vapor present a minimum for the same couple of ($T_{\mathrm{MOL}},\theta_{\mathrm{MOL}}$). Therefore our hypothesis of a single layer containing both elements is validated. Our best fit values for the angular diameter and the temperature of the MOLsphere are close to those derived by Perrin \etal\ (\cite{Perrin2004}) which were $T_{\mathrm{MOL}} = 2055 \pm 55 \, \mathrm{K}$ and $\theta_{\mathrm{MOL}} = 55.78 \pm 0.04 \, \mathrm{mas}$.

	\section{Conclusion}
	
		Our analysis of the AMBER visibilities in the continuum allowed us to compute new uniform disk and limb-darkened disk angular diameters. Our absorption line analysis led us to new values for the MOLsphere parameters, consistent with the results from Perrin \etal\ (\cite{Perrin2004}) and Ohnaka \etal\ (\cite{Ohnaka2011}).\\
		
		However, our analysis here only includes data from the first two CO overtone band heads : it must be extended to the other bands, that may probe a deeper molecular layer, closer to the star's photosphere. Finally we will conclude this analysis by exploiting the closure phase signal. This work is currently undergoing and will allow us to search for asymmetries in the light distribution at the surface of the star.  
	
	\begin{acknowledgements}
	This research has made use of the  \texttt{AMBER data reduction package} of the Jean-Marie Mariotti Center\footnote{Available at \url{http://www.jmmc.fr/amberdrs}}
	\end{acknowledgements}

\end{document}